\documentclass[prl,aps,showpacs,twocolumn,groupedaddress]{revtex4}

\usepackage{amsmath,amsfonts,amssymb}
\usepackage{wrapfig}
\usepackage{graphicx}
\usepackage{bbm}
\usepackage{graphics}

\def \beq {\begin{equation}}
\def \eeq {\end{equation}}
\def \ba {\begin{eqnarray}}
\def \ea {\end{eqnarray}}
\newcommand{\upp}{\hspace{-0.2 pt}\uparrow}
\newcommand{\downn}{\hspace{-0.2 pt}\downarrow}

\newcommand{\ketbrad}[1]{|#1\rangle\!\langle #1|}
\newcommand{\ketbra}[2]{|#1\rangle\!\langle #2|}

\newcommand{\mean}[1]{\langle#1\rangle}

\def\ket#1{\left| #1\right>}
\def\bra#1{\left< #1\right|}
\renewcommand{\section}[1]{\emph{#1}: }
\renewcommand{\subsection}[1]{\emph{#1} --- }

\begin{document}

\title{Electrically-protected resonant exchange qubits in triple quantum dots}
\author{J. M. Taylor$^1$, V. Srinivasa$^1$, J. Medford$^2$}
\affiliation{$^1$Joint Quantum Institute/National Institute of Standards and Technology, College Park, Maryland 20742, USA}
\affiliation{$^2$Department of Physics, Harvard University, Cambridge, Massachusetts 02138, USA}
\begin{abstract}
We present a modulated microwave approach for quantum computing with qubits comprising three spins in a triple quantum dot.  This approach includes single- and two-qubit gates that are protected against low-frequency electrical noise, due to an operating point with a narrowband response to high frequency electric fields.  Furthermore, existing double quantum dot advances, including robust preparation and measurement via spin-to-charge conversion, are immediately applicable to the new qubit.  Finally, the electric dipole terms implicit in the high frequency coupling enable strong coupling with superconducting microwave resonators, leading to more robust two-qubit gates.
\end{abstract}
\maketitle

Spins in quantum dots as an architecture for quantum information processing require some combination of electric and magnetic field control at the nanometer scale~\cite{loss98}.  While the intrinsic coherence properties of the spins can be remarkable, the need for such control inevitably couples the qubit degree of freedom to low-frequency electric or magnetic noise~\cite{schliemann01,erlingsson02,barrett02,coish05,taylor.2005.177--183,taylor.2007.035315,valente.2010.125302}.  Approaches that mitigate this coupling, via dynamical decoupling or composite pulse sequences, all require rapid `pulsed gate' control either for individual qubits or for two-qubit gates, which in turn requires wide bandwidths for the control electronics.  While this has led to a variety of advances in the field, paradoxically it also leads to the use of quantum bits as sensors, rather than as protected devices~\cite{dial.2013.146804}.

Instead, we suggest that the use of so-called exchange-only qubits~\cite{divincenzo00,lidar00}, comprising three spins in a triple quantum dot~\cite{vidan05,hawrylak2005voltage,gaudreau06,korkusinski.2007.115301,rogge2009three,hsieh2012physics} and implemented experimentally~\cite{laird.2010.,takakura2010triple,gaudreau.2012.54--58,medford.2013.}, provide an opportunity for protection against low-frequency control noise in analogy to advances in superconducting devices~\cite{houck.2009.105--115}.  In particular, by having exchange couplings always on, a regime with no low-frequency field response and a narrowband, resonant response becomes accessible.  We denote this the resonant exchange (RX) qubit, and refer the reader to the concurrent Ref.~\cite{medford.2013b} for an experimental demonstration of these ideas.  Furthermore, our approach has a protected two-qubit interaction via exchange~\cite{srinivasa.2009.,doherty.2013.} or via resonant dipole-dipole interactions. 
As coupling between qubits relies on electric fields rather than tunneling, devices could be implemented in a wide variety of potential materials such as two dimensional electron gas and nanowire depletion dots.  Finally, we show that the dipolar nature of the RX qubit also enables strong coupling with high quality factor microwave cavities.

\begin{figure}
\includegraphics[width=3.4in]{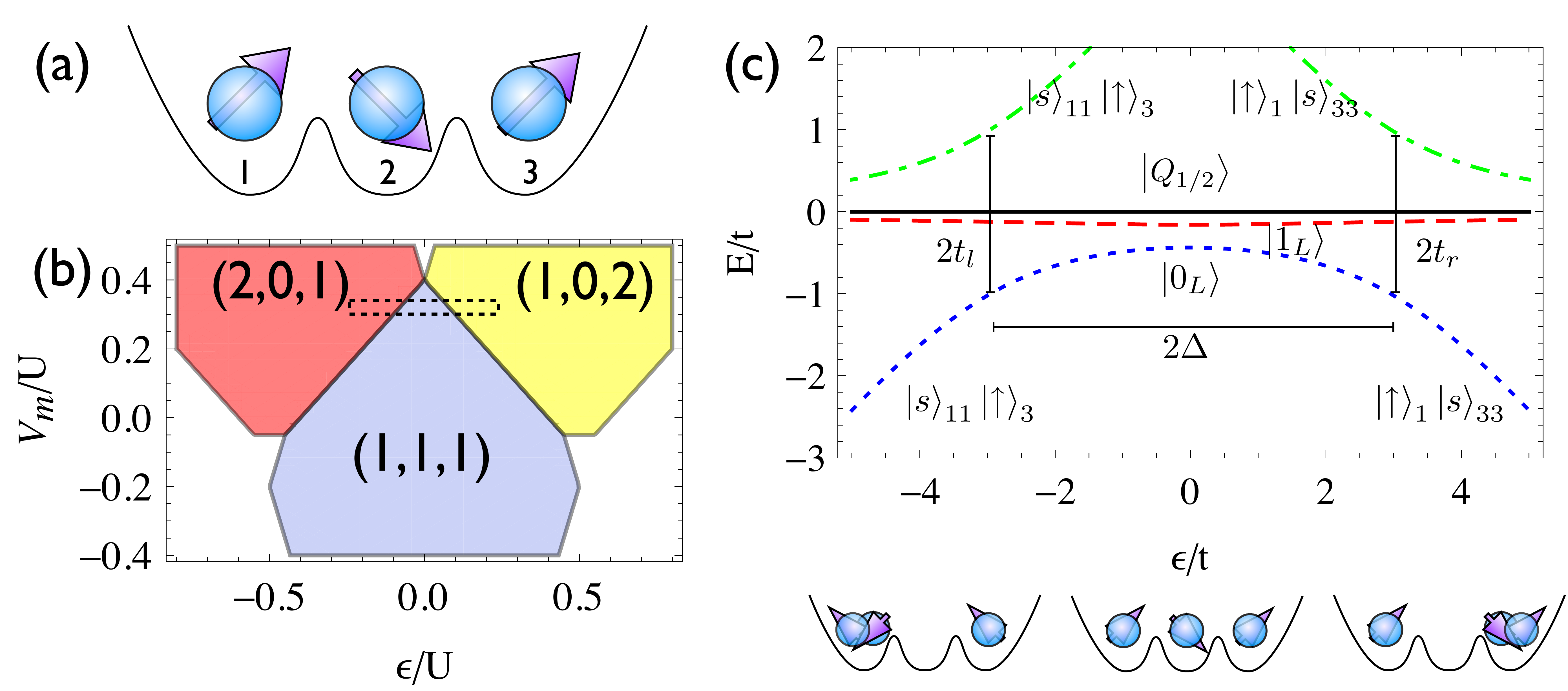}
\caption{
\label{f:energies}
(a) Schematic of a triple quantum dot with one electron per dot -- the (1,1,1) charge configuration.
(b) Charge stability diagram for a lateral triple dot with $U_c = 0.3 U$,  $V_{\rm tot} = 2.6 U$, $t_{l}=t_{r}=0$ as a function of
$\epsilon/U, V_m/U$ where $U \sim 1-10$ meV.  As $V_m$ increases the width of the (201)-(111)-(102) region decreases, and the dashed box indicates the regime of interest for (c); other charging numbers are not shown.
(c) Energy levels at high external magnetic field ($g^* \mu_B B_{\rm ext} \gg t$) for the $m_s = 1/2$ subspace as a function of $\epsilon$ for fixed $\Delta = 3 t$ in units of $t = t_l = t_r$.
}
\end{figure}

The few electron regime of interest for our triple dot system we describe by the Hubbard model~\cite{korkusinski.2007.115301}
\beq
H_{\rm hub} = \sum_i \frac{U}{2} n_i(n_i-1) - V_i n_i + \sum_{\mean{i,j}} U_c n_i n_j - \frac{t_{ij}}{\sqrt{2}}  c_{i,\sigma}^\dag c_{j,\sigma} 
\eeq
where $U$ is the individual dot charging energy, $U_c$ is the cross-charging energy, $V_i$ is the local potential set by applied gate voltages on dot $i$, $t_{ij}$ is the tunneling between dots $i$ and $j$, and $c^\dag_{i\sigma}$ is the creation operator for an electron on dot $i$ with spin $\sigma$. For simplicity, we assume a linear array and set $t_{12} = t_l, t_{23} = t_r, t_{13} = 0$, and have defined tunneling such that singlet-singlet tunneling has a rate $t_{l,r}$.  An appropriate choice of gate voltages makes the charge stability region (1,1,1) relatively narrow (Fig.~\ref{f:energies}a,b), where ($n_1,n_2,n_3$) indicates charge occupation numbers for each dot.  
Specifically, defining $\epsilon = (V_3 - V_1)/2$ and $V_m =  (V_1 + V_3)/2 - V_2$, we find the (1,0,2)-(1,1,1) [(2,0,1)-(1,1,1)] charge transitions occur at $\epsilon = \pm \Delta \equiv \pm (U - 2 U_c - V_m)$, which also defines the width $2 \Delta$ of the (1,1,1) region.  We see that by making $V_m$ sufficiently large, this window width can be made arbitrarily small.  At the same time, triple dot configurations with a total charge number different than three can remain detuned by keeping the independent parameter $V_{\rm tot}  = \sum_i V_i$ in an appropriate range.

Restricting our discussion to the reduced model with only (2,0,1), (1,1,1), and (1,0,2),
for large positive or negative $\epsilon$, the charge avoided crossing behaves in a manner analogous to a double quantum dot~\cite{taylor.2007.035315,laird.2010.}, with the third charge (and spin) largely decoupled.  Near these avoided crossings (shown in Fig.~\ref{f:energies}c), techniques from two-spin quantum bits, including singlet preparation and measurement~\cite{taylor05}, become available and provide a means of initializing and measuring the states of the total spin $S=1/2$ manifold.  Recently the necessary adiabatic mapping between the (2,0,1) singlet state and the lowest energy $S=1/2$ state in (1,1,1) has been shown~\cite{laird.2010.,medford.2013.}.

Moving to the qubit itself, we consider small $|\epsilon|$.  In this regime, all three spins are undergoing exchange via virtual tunneling to the doubly occupied singlet states $\ket{s_{1,m_s}} \equiv  \ket{s}_{11}\ket{m_s}_{3}, \ket{s_{3,m_s}} \equiv \ket{m_s}_{1}\ket{s}_{33}$, with $m_s=\pm 1/2$ and $\ket{s}_{ij}$ denotes a singlet of spin for electrons in dots $i$ and $j$.  A Schrieffer-Wolff transformation, defined by $\exp(\lambda A) H \exp(-\lambda A)$ such that the transformed Hamiltonian is block diagonal in charge to $\mathcal{O}(\lambda^2)$ with the assumption $t_{l,r} \propto \lambda$, yields a coupled Heisenberg  model:
\beq
H_{\rm Heis} = J_{l} S_1 \cdot S_2 + J_{r} S_2 \cdot S_3 
+ \mathcal{O}(t \xi^3)
\eeq  
where $J_{l} = \frac{t_l^2}{\Delta + \epsilon}, J_{r} = \frac{t_r^2}{\Delta - \epsilon}$ are left-center and right-center exchange.  We further define $\xi = t/\Delta$ as our charge admixture parameter, determining when corrections to the Heisenberg model may become important.

Under the application of a large external magnetic field (assumed $g^* \mu_B B_{\rm ext} \gg t$), the Zeeman sub-levels of the Heisenberg chain split, and we are left with subspaces defined by the additional quantum numbers for the spins.  In contrast to the case of two spins, where the only additional number is the total spin $S$, here we have both the total spin and a symmetric group quantum number.  It is in this symmetric group (permutation) sector that our quantum bit will be defined, following earlier work~\cite{divincenzo00}, and we label this regime of operation the RX regime.  The logical subspace with $S=1/2$ is
\begin{eqnarray}
\ket{1} &= & \ket{s}_{13} \ket{\upp}_2 \ ,\\
 \ket{0} & = & \sqrt{1/3} \ket{t_0}_{13} \ket{\upp}_2 - \sqrt{2/3} \ket{t_+}_{13} \ket{\downn}_2\ .
 \end{eqnarray}  
$\ket{t_{0,\pm}}$ are the three spin triplet states.  The third state, $\ket{Q_{1/2}} = \sqrt{2/3} \ket{t_0}_{13} \ket{\upp}_2 + \sqrt{1/3} \ket{t_+}_{13} \ket{\downn}_2$ is a state with total $S = 3/2$ and has zero energy, as shown in Fig.~\ref{f:energies}c.     We will work in this subspace $P$ for the remainder of this Letter.  Defining $J = \frac{J_{l}+J_{r}}{2},j = \frac{J_{l} - J_{r}}{2}$,
\beq
PH_{\rm Heis}P = -\frac{3 J}{2} \ketbrad{0} - \frac{J}{2} \ketbrad{1} -\sqrt{3/4} j \left(\ketbra{0}{1} +\ketbra{1}{0}\right)\ .
\eeq

\section{Single-qubit gates} The RX qubit is designed to operate at the point in parameter space where the fixed electric fields yield qubit a with no low-frequency response to additional fields (no DC dipole), but with a narrow-band response to an applied AC field at the qubit frequency (an AC dipole).  
In the context of our model, variations of $\epsilon \propto e \sum_i x_i$ (with $x_i$ the position of dot $i$ along the inter-dot axis) are equivalent to looking at the electric dipole response of the qubit system (Fig.~\ref{f:2q}a).  Our ideal operation point -- protected against low-frequency noise by the energy gap $\sim J$ -- occurs when the energy quantization axis in the $\ket{0},\ket{1}$ subspace for a fixed $\epsilon_0$ is perpendicular to the perturbation of the system due to a small variation $\epsilon = \epsilon_0 + F$.  For small tunneling asymmetry $t_{r(l)} = (1 \pm y) t$, we have $\epsilon_0 = -\frac{8 \Delta}{5} y$.
Variations of $\epsilon$ around $\epsilon_0$ lead to variations of a perturbation orthogonal to the quantization axis, providing a direct dipole moment between ground and excited states as indicated by the virtual process in Fig.~\ref{f:2q}a.  We write the Hamiltonian using Pauli matrices such that its eigenenergies are quadratic in $F$:
\beq
H_{\rm RX} = \frac{\hbar \omega}{2} \sigma_z + F \eta \sigma_x
\eeq
with $\hbar \omega = \sqrt{J^2 + 3 j^2} \approx t \xi (1 + \frac{3 y^2}{5})$ the energy difference between the two energy eigenstates and $\eta = \sqrt{(\partial_\epsilon J)^2+3 (\partial_\epsilon j)^2}|_{\epsilon = \epsilon_0} \approx  2 \sqrt{3} \xi^2 (1+ \frac{63 y^2}{25})$ an effective matrix element coupling $F$ to the qubit degree of freedom.  

Rabi nutation becomes accessible if one drives $\epsilon$ sinusoidally, i.e., $F(\tau) = f(\tau) \cos(\nu \tau +\phi)$ where $\phi$ is an adjustable phase and $f(\tau)$ varies slowly compared to the gap $\omega$.  Near resonance (small $\delta = \omega - \nu$), we can work in a rotating frame and rotating wave approximation, and have a quantum bit nutation with Rabi frequency 
\beq
\Omega(\tau) \approx \frac{f(\tau)}{\hbar} \sqrt{3} \xi^2 \ .
\eeq
Subsequent Rabi nutations, with differing $\phi$, correspond to rotations about various axes along the equator of the Bloch sphere defined in $H_{\rm rot}$.

The key feature of the RX qubit is a reduction in response to low-frequency noise, due to the gap $\omega$.
However, a secondary benefit arises from purely practical concerns: the necessary bandwidth for performing gates via modulated microwaves is on the order of $1/\tau_{\rm gate}$ around the center frequency $\omega$.  This allows for filtering of classical noise in other frequencies to a high order.  However, the adiabatic gate control necessary in the current scheme for preparation and measurement requires a second bandpass for control at a frequency much lower than $\omega$.  Alternatively, as we show below, preparation and readout can be accomplished using circuit QED~\cite{blais04,wallraff04} coupling of the qubit levels to a microwave cavity, in direct analogy to the transmon superconducting qubit~\cite{reed.2010.173601,
mallet.2009.791--795}.

\section{Two-qubit gates}
\begin{figure}
\includegraphics[width=3.2in]{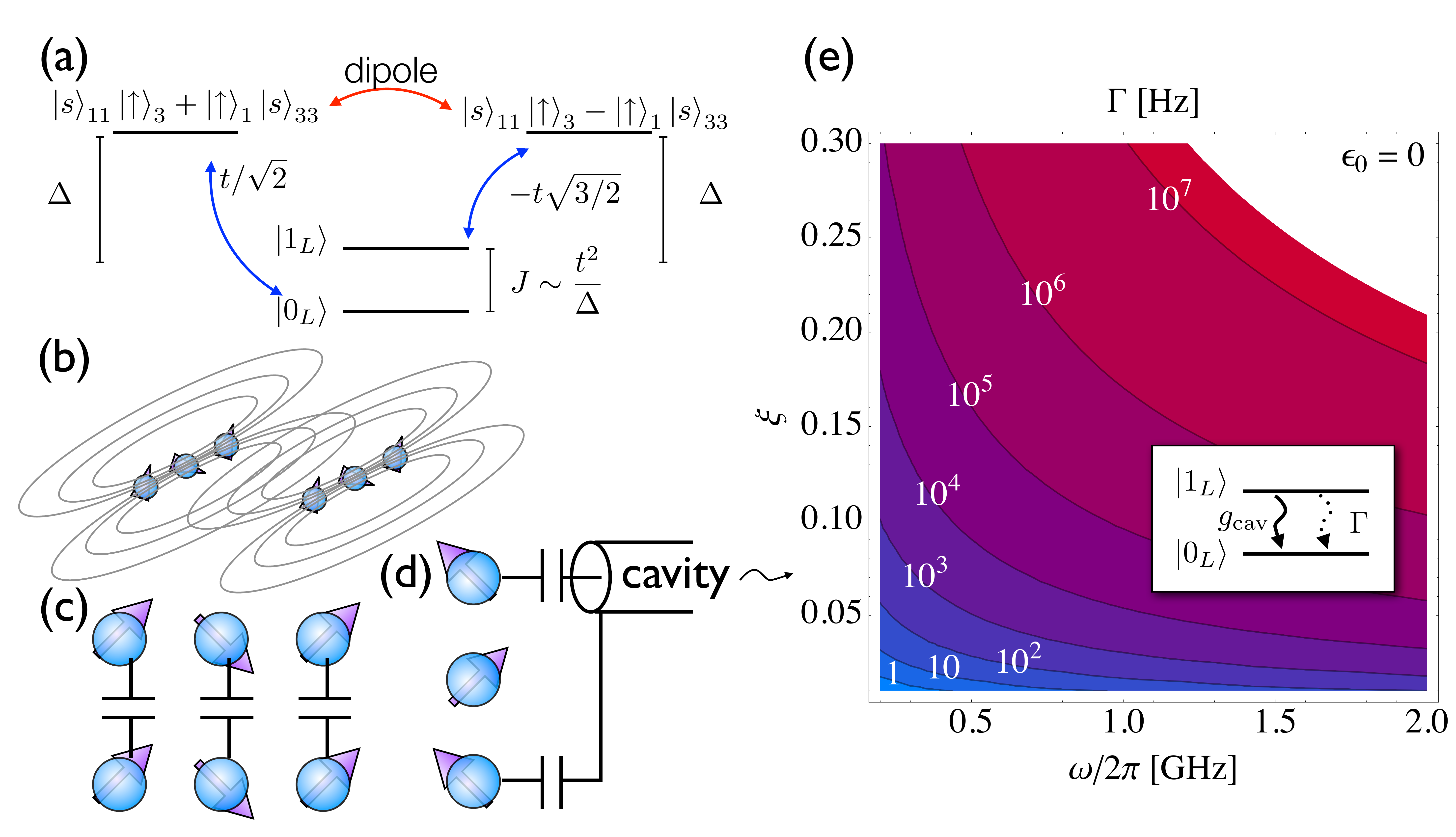}
\caption{
\label{f:2q}
(a) Virtual process that leads to an electric dipole element between the two logical states, where tunneling couples to the symmetric (anti-symmetric) combination of excited charge states, which are in turn coupled by electric field.
(b-d) Dipole-dipole coupling, near-field multipole coupling, and coupling to a superconducting transmission line resonator.
(e) Contour plot of  phonon-induced relaxation $\Gamma$ between logical states
at the operation point $\epsilon_{0}=0,$ as a function of the qubit frequency $\omega$
and the charge admixture parameter $\xi=t/\Delta.$ Quantities used
to calculate $\Gamma$ include the Gaussian width parameter $\sigma=20\ \mbox{nm}$,
$a=260\ \mbox{nm},$ and phonon parameters relevant for GaAs quantum
dots \cite{Stano2006PRL}, including $\rho_{0}=5.3\times10^{3}\ \mbox{kg/m}^{3},$
$c_{l}=5.3\times10^{3}\ \mathrm{m}/\mathrm{s},$ $c_{t}=2.5\times10^{3}\ \mathrm{m}/\mathrm{s},$
$\beta_{l}=7.0\ \mbox{eV},$ and $\Xi=1.4\times10^{9}\ \mbox{eV/m}$.  Inset shows how electric dipole coupling to the cavity also leads to phonon-based decay.
}
\end{figure}

In addition, this so-called `sweet spot' provides a several mechanisms for coupling RX qubits.  In particular, two RX qubits in proximity will have a multipole-multipole interaction due to the coupling of the dipole to the electric field.  At large distances $R \gg a$ between two RX qubits with relative angle $\theta$ and dipole moments $\vec{d}_{a(b)}$, they interact via dipole-dipole terms (Fig.~\ref{f:2q}a,b)
\beq
V_{dd} = \frac{\vec{d}_a \cdot \vec{d}_b (1- 3 \cos^2 \theta)}{4 \pi \varepsilon R^3} \approx \hbar g_{dd} (\sigma_+^{(a)} \sigma_-^{(b)} + \textrm{H.c.})
\eeq
with an overall strength $\hbar g_{dd} \sim \frac{3 e^2 a^2}{4 \pi \varepsilon R^3}\xi^4$, where $a$ is the left-right dot distance, $\varepsilon$ is the dielectric constant of the substrate, and non-secular terms have been neglected. This enables direct `swap'-type gates between the two qubits.  Extensions of CORPSE pulse sequences allows for such gates to be integrated with homogeneous echo sequences~\cite{marcos.2010.210501}, providing potential high-fidelity operation even for slow gates as implicit in the overall $\xi^4$ performance.  For qubits with resonance frequencies differing by $\delta$, driving one qubit with a Rabi frequency $\Omega = \delta$ creates, in the double rotating frame, a resonant interaction, in direct analogy to Hartman-Hahn double resonance.

While a dipolar approximation yields a straightforward understanding of the coupling mechanism, the more general result will require inclusion of higher order terms, as $a \sim R$.  Fortunately, a convenient approximation remains by expanding the interaction between the two RX qubits via their capacitive interaction, where we can include an explicit coupling between adjacent RX qubits of the form $U_{ab} \sum_{i} n_{a,i} n_{b,i}$ (Fig.~\ref{f:2q}c).  This is a direct extension of efforts to couple double quantum dots capacitively~\cite{taylor.2005.177--183,weperen.2011.030506,shulman.2012.202--205}.  In the transformed frame, 
\beq
n_{a,1(3)} -1 \approx \frac{t^2_{l(r)}}{4 (\Delta \pm \epsilon)^2} \left[ -\sigma_z^{(a)} \pm \sqrt{3} \sigma_x^{(a)} \right]
\eeq
Written in the logical basis with $t_l = t_r = t$, we have
\beq
V_{mm} = U_{ab} \frac{\xi^4}{2} \left( \sigma_z^{(a)} \sigma_z^{(b)} + 3 \sigma_x^{(a)} \sigma_x^{(a)} \right)
\eeq
with non-secular terms such as $\sigma_z^{(a)} \sigma_x^{(a)}$ removed by symmetry.  Going beyond the dipole expansion has added a $z-z$ coupling term, which can enable gates even when the qubits are at different resonant frequencies, as is commonly dealt with in NMR~\cite{vandersypen04}.

Finally, an additional mechanism for two-qubit gates is the direct coupling of a RX qubit to a high quality factor superconducting cavity (Fig.~\ref{f:2q}d), as has recently been achieved for double quantum dots~\cite{petersson.2012.380--383,frey.2012.046807}, where $g_{\rm chg}/\omega \gtrsim 10^{-3}$ was demonstrated.  We can estimate the vacuum Rabi coupling of a dipole of size $e a$ ($a \sim 260$ nm is the size of the triple dot system) to a transmission-line circuit QED cavity with transverse mode area $A$ as $g_{\rm chg} \approx \omega \frac{\alpha^{1/2} a}{ \sqrt{A}}$ with $\alpha$ the fine structure constant.  The corresponding coupling of the RX qubit is $g_{\rm cav} = g_{\rm chg} \sqrt{3} \xi^2$.
For $\sqrt{A} \sim 3\ \mu$m, we have $g_{\rm cav}/\omega \sim 7 \times 10^{-3} \xi^2$.  Thus for $\xi \gtrsim 0.03$ and for systems with high cavity $Q \gtrsim 10^6$ and long qubit $T_2 \gtrsim 20\ \mu$s~\cite{medford.2013b}, the strong coupling regime becomes accessible.

\section{Corrections from spin-orbit and nuclear spins}
One potential complication is the addition of spin-orbit coupling.  However, if we neglect spin-flip terms in the spin-orbit coupling due to the large applied external field, the resulting spin-conserving terms correspond only to a redefinition of the singlet states.  Specifically, including as an addition to tunneling terms $-i \lambda_{l}\sum_{\sigma} \frac{\sigma}{\sqrt{2}} c_{2\sigma}^\dag c_{1\sigma}$ and similarly for $r$, the qubit subspace is defined by the two non-orthogonal basis vectors $\ket{\tilde{0}} \propto (t_l - i \lambda_l) \ket{\downn \upp \upp} - (t_l + i \lambda_l) \ket{\upp \downn \upp}, \ket{\tilde{1}} \propto (t_r - i \lambda_r) \ket{\upp \downn \upp} -  (t_r + i \lambda_r) \ket{\upp \upp \downn}$.  Thus spin-conserving spin-orbit serves only to renormalize the effective tunneling coefficients.

While spin-orbit coupling is mostly accounted for, nuclear spins and other magnetic gradient fields change can lead to additional noise~\cite{ladd.2012.125408}.  In particular, for Zeeman terms $B_i$ in each dot aligned with the external homogeneous field (due to nuclear gradients or to other gradients), we have in the $\ket{0},\ket{1},\ket{Q_{1/2}}$ basis
\beq
V_{\rm nuc} = \frac{g^* \mu_B m_s}{2} \left( \begin{array}{ccc}
-B_Q & B_D & \sqrt{2} B_D \\
B_D & B_Q & -\sqrt{2} B_Q\\
\sqrt{2} B_D & -\sqrt{2} B_Q & 0 
\end{array} \right)
\eeq
with $B_Q = \frac{1}{3} (B_1 - 2 B_2 + B_3)$ and $B_D = (B_3 - B_1)/\sqrt{3}$ the quadrupolar and dipolar contributions from the gradient.  We remark that the nuclear effects, both on the qubit frequency $\omega$ and on the coupling between the two qubit states and leakage state may be suppressed for triple dot devices with a non-trivial tunneling between dots 1 and 3, as recognized in molecular magnet studies~\cite{troiani.2012.,georgeot.2010.200502,trif.2008.217201}.

Fortunately, while gradient terms change the Hamiltonian, they may suppressed by two different effects.  First, the coupling between qubit states (and to the leakage space) is suppressed by the large energy scale $\omega$ set by exchange couplings.  In essence, exchange averages over nuclear gradients, leading only to higher order effects from all terms except the diagonal terms in the Hamiltonian (which go as $B_Q$ for the symmetric regime).  Second, the $B_Q$ terms are slowly varying with root-mean-square expectation of $\sqrt{2/3} B_{\rm nuc}$ where $B_{\rm nuc}$ is the average size of the nuclear field in a single quantum dot.  As such, they may be suppressed by use of standard spin-echo due to their low-frequency character.  

\section{Coupling to phonons}
Phonons provide an intrinsic
source of electric field fluctuations that couple to the electric
dipole moment of the qubit via the electron-phonon interaction, causing
relaxation \cite{Hanson2007RMP,Mehl2012}. This interaction has the
form \cite{Mahan2000} 
\begin{eqnarray}
H_{\mathrm{ep}} & = & \sum_{\mu,\mathbf{k}} M_{k} \sqrt{\frac{\hbar}{2\rho_{0}V_{0}c_{\mu}k}}\left(k\beta_{l}\delta_{\mu,l}-i\Xi\right)\left(a_{\mu,\mathbf{k}}+a_{\mu,-\mathbf{k}}^{\dagger}\right) \nonumber \\ 
M_{k} & \equiv & \sum_{i,j=1}^{3}\sum_{\sigma}\bra{i}e^{i\mathbf{k}\cdot\mathbf{r}}\ket{j}c_{i,\sigma}^{\text{\dag}}c_{j,\sigma} \nonumber \label{eq:mk}
\end{eqnarray}
where $M_k$
is the factor which depends on electronic degrees of freedom, and the constants are:
mass density $\rho_{0}$,  volume $V_{0}$,  phonon speeds $c_{\mu}$,
 deformation potential $\beta_{l}$ for the longitudinal $\left(l\right)$
mode, and piezoelectric constant $\Xi$. The operator $a_{\mu,\mathbf{k}}^{\dagger}$
creates 
a phonon with wavevector
$\mathbf{k}$, energy $\varepsilon_{\mathrm{ph}}=\hbar c_{\mu}k,$
and polarization $\mu$, and $\delta_{\mu,l}$ is the Kronecker delta
function.

The rate $\Gamma=T_{1}^{-1}$ of qubit relaxation due to $H_{\mathrm{ep}}$
is given by Fermi's golden rule as$ $ $\Gamma\sim\left|\left\langle 0\right|H_{\mathrm{ep}}\left|1\right\rangle \right|^{2}\rho\left(\omega\right)$.
Here, $\rho\left(\omega\right)$ is the phonon density of states evaluated
at the exchange gap $\omega$ between the states $\left|0\right\rangle $
and $\left|1\right\rangle $ that determines the energy of the emitted
phonon. The matrix element $\bra{0}M_{k}\ket{1}$ is evaluated by
defining Gaussian wavefunctions $\psi_{i}\left(\mathbf{r}\right)\equiv\left\langle \mathbf{r}\right|\left.i\right\rangle $
which are shifted along the dot axis by $-a/2,$ $0,$ and $a/2$
for $i=1,2,$ and $3,$ respectively, and by changing to the qubit
basis via the same Schrieffer-Wolff and diagonalization transformations
used to determine $H_{\mathrm{eff}}$. 
A simple model suggests $\Gamma \sim \frac{\omega^3}{\nu_\Gamma^2} \xi^4$, given the piezoelectric coupling domination at low frequencies and the dipole moment of the qubit, which is consistent with the more detailed prediction (Fig.~\ref{f:2q}e) fit with $\nu_\Gamma \sim 2 \pi \times 0.9$ GHz.  The simple and detailed models agree over the range $\frac{\omega}{2\pi} = 0-2$ GHz and $\xi = 0-0.3$.

\section{Qubit performance}
We now seek optimal parameters for performance.  We assume $t_l = t_r = t$ (setting $\epsilon_0 = 0$) and use $\xi$ and $\omega = t^2/\Delta$ to rewrite all terms with $t$ or $\Delta$.
Nuclear spin-induced $T_2^*$ is due to fluctuations in $B_Q$.  Variations in $\epsilon$ are limited the width of our narrowband filter; thus $\Omega \ll \omega$.  A single qubit $\pi$ rotation has an estimated infidelity $I_1 \sim \pi^2/(\Omega T_2^*)^2 + \pi/(\Omega T_2) + (\Omega/(\sqrt{3} \omega))^2$.  For long $T_1$ times, this yields $I_1 \sim \frac{\sqrt{8/3} \pi }{T_2^* \omega}$, though for sufficiently large $\omega$, $T_2$ becomes important.

However, two-qubit gates are more limited by $T_2$ than by $T_2^*$, as we can apply refocusing pulses during the two-qubit gate.  In that case, we seek to balance the gate time against $T_2$.  As $T_2$ may be limited by phonon relaxation at appreciable values of $\omega$, we consider the $T_1$ limited scenario, with an estimate of the infidelity for an entangling gate $I_2 \sim \frac{2 \pi}{U_{ab} \xi^4 T_1} = \frac{2 \pi \omega^3}{U_{ab} \nu_\Gamma^2}$.  Also requiring $\omega \gg 1/T_2^*$ for single-qubit gates, an optimal qubit frequency is set by $\omega \sim \sqrt{\nu_\Gamma}(U_{ab}/T_2^*)^{1/4}$.  For $U_{ab} \sim 2 \pi \times 30$ GHz and $1/T_2^* \sim 2 \pi \times 10 $ MHz, consistent with current GaAs devices~\cite{shulman.2012.202--205}, the optimal $\omega \sim 2 \pi \times 0.7$ GHz and $I_2 \lesssim 0.08$.  Some improvement can occur by reduction of the nuclear field, improvement in capacitive coupling, or phonon band-gap engineering.  However, the most dramatic increase of fidelity occurs in the presence of a resonant superconducting cavity.  Then the product of cavity coupling and relaxation, $g_{\rm cav} T_1$, increases with smaller $\xi$, leading to gates that may be limited by $T_2$ processes.

Additional benefits to the resonance exchange approach may be achieved by exploring other coupling and control mechanisms, such as direct tunnel coupling between adjacent RX qubits~\cite{doherty.2013.}.

The authors acknowledge helpful discussions with C. M. Marcus, D. DiVincenzo, T. Ladd, M. Gyure, D. Lidar, D. Loss, A. Yacoby, J. Beil, and E. Rashba.  Work was supported by DARPA MTO and by the NSF funded Physics Frontier Center at the JQI.

\end{document}